\documentclass[preprint,aps,a4paper]{revtex4-1}

\usepackage{mathrsfs}
\usepackage{graphicx}
\usepackage{multirow}
\usepackage{makecell}
\usepackage{booktabs}
\usepackage{lineno}

\begin{document}
    \title{Conservation of the Particle-Hole Symmetry in the Pseudogap State in Optimally-Doped $\mathbf{Bi_{2}Sr_{2}CuO_{6+\delta}}$ Superconductor}	
		
	\author{Hongtao Yan$^{1,2}$, Qiang Gao$^{1}$, Chunyao Song$^{1,2}$, Chaohui Yin$^{1,2}$, Yiwen Chen$^{1,2}$, Fengfeng Zhang$^{3}$, Feng Yang$^{3}$, Shenjin Zhang$^{3}$, Qinjun Peng$^{3}$, Guodong Liu$^{1,2,4}$, Lin Zhao$^{1,2,4}$, Zuyan Xu$^{3}$ and X. J. Zhou$^{1,2,4,5,*}$}
		
	\affiliation{
		\\$^{1}$National Lab for Superconductivity, Beijing National Laboratory for Condensed Matter Physics, Institute of Physics, Chinese Academy of Sciences, Beijing 100190, China
		\\$^{2}$University of Chinese Academy of Sciences, Beijing 100049, China
		\\$^{3}$Technical Institute of Physics and Chemistry, Chinese Academy of Sciences, Beijing 100190, China
		\\$^{4}$Songshan Lake Materials Laboratory, Dongguan, Guangdong 523808, China
		\\$^{5}$Beijing Academy of Quantum Information Sciences, Beijing 100193, China
		\\$^{*}$Corresponding author: XJZhou@iphy.ac.cn
	}
	
	\date{\today}
	
	\maketitle
	
	\newpage
	
	{\bf The pseudogap state is one of the most enigmatic characteristics in the anomalous normal state properties of the high temperature cuprate superconductors. A central issue is to reveal whether there is a symmetry breaking and which symmetries are broken across the pseudogap transition. By performing high resolution laser-based angle-resolved photoemission measurements on the optimally-doped Bi$_{2}$Sr$_{1.6}$La$_{0.4}$CuO$_{6+\delta}$ superconductor, we report the observations of the particle-hole symmetry conservation in both the superconducting state and the pseudogap state along the entire Fermi surface. These results provide key insights in understanding the nature of the pseudogap and its relation with high temperature superconductivity.	
	}
	
	\vspace{3mm}
	
	High temperature cuprate superconductors exhibit a number of anomalous properties in the normal state. One prominent anomaly is the observation of the pseudogap that opens above the superconducting transition temperature $T_{c}$ but below the characteristic temperature $T^{*}$ that is usually defined as a pseudogap temperature\cite{BStatt1999TTimusk,AKapitulnik1996AGLoeser,JGiapintzakis1996HDing}. Revealing the nature of the pseudogap and its relation with superconductivity have been central issues in understanding the mechanism of high temperature superconductivity. It remains under debate whether the pseudogap is associated with the pre-formed pairing\cite{SAKivelson1995VJEmery} or some competing orders\cite{JRSchrieffer1990AKampf,CNayak2001SChakravarty,DHLee2006JXLi,ZXShen2010MHashimoto,ZXShen2011RHHe,PALee2014PALee,ZXShen2019SDChen}. Since the pre-formed pairing usually displays particle-hole symmetry while the competing orders may not, the examination of the particle-hole symmetry is crucial to understanding the nature of the pseudogap. In the underdoped $\mathrm{Bi_{2}Sr_{2}CaCu_{2}O_{8+\delta}}$ (Bi2212) superconductor, it is reported from the angle-resolved photoemission (ARPES) measurements that, in the pseudogap state, the particle-hole symmetry breaks near the nodal region but is conserved near the antinodal region\cite{GDGu2008HBYang}. In the optimally-doped $\mathrm{Bi_{2}Sr_{2}CuO_{6+\delta}}$ (Bi2201) superconductor, dramatic electronic structure change is observed across $T^{*}$ and $T_{c}$ over a wide momentum space, suggesting a phase transition across $T^{*}$ and the breaking of the particle-hole symmetry and the spatial symmetry in the pseudogap state\cite{ZXShen2010MHashimoto,ZXShen2011RHHe}. These unusual phenomena are interpreted in terms of the pair density wave formation in cuprate superconductors\cite{PALee2014PALee}. Considering the importance of the particle-hole symmetry in understanding the nature of the pseudogap, it is crucial to establish whether the observed phenomena in Bi2212\cite{GDGu2008HBYang} and Bi2201\cite{ZXShen2010MHashimoto,ZXShen2011RHHe} are intrinsic and universal in the pseudogap state of the cuprate superconductors.		
	
	In this paper, we report the observations of the particle-hole symmetry conservation in both the superconducting state and the pseudogap state by performing high resolution laser-based ARPES measurements on the optimally-doped Bi$_{2}$Sr$_{1.6}$La$_{0.4}$CuO$_{6+\delta}$ (La-Bi2201) superconductor. The Fermi surface topology and the band structures exhibit little change with temperature across the pseudogap temperature $T^{*}$. The particle-hole symmetry is observed along the entire Fermi surface both in the pseudogap state and in the superconducting state. These results provide key insights in understanding the nature of the pseudogap and its relation with high temperature superconductivity.
			
	The ARPES measurements were performed by using our lab-based laser ARPES system equipped with the 10.897\,eV vacuum-ultra-violet (VUV) laser and a angle-resolved time-of-flight electron energy analyzer (ARToF) which can simultaneously detect the two-dimensional momentum space\cite{XJZhou2008GDLiu,WTZhang2018XJZhou}. The energy resolution was set at $\sim$1\,meV and the angular resolution is $\sim$0.1\,$^\circ$, corresponding to a momentum resolution of $\sim$0.0023\,\AA$^{-1}$ at the photon energy of 10.897\,eV. High quality single crystals of the optimally-doped La-Bi2201 were grown by the traveling solvent floating zone method. The samples were post annealed in the flowing oxygen to adjust the hole concentration and make the samples uniform\cite{XJZhou2009JQMeng2}. For convenience, we use Opt32K to represent the optimally-doped Bi2201 sample with a $T_{c}$ of 32\,K. The pseudogap temperature $T^{*}$ is $\sim$150\,K as determined from ARPES and NMR measurements\cite{CTLin2005GQZheng,VMadhavan2008JHMa}. The sample was cleaved $\mathit{in\,situ}$ at 20\,K and measured in vacuum with a base pressure better than 5$\times$10$^{-11}$\,Torr. The Fermi level is referenced by measuring on a clean polycrystalline gold that is electrically connected to the sample and also by the ARPES data along the nodal direction which are known to have zero superconducting gap.    
	
    Figure 1 shows the Fermi surface mappings of the Opt32K Bi2201 sample measured at different temperatures across both the superconducting transition temperature $T_{c}$ of 32\,K and the pseudogap temperature $T^{*}$ of $\sim$150\,K. It consists of two separate measurements: one is centered around the nodal region (Fig. 1a1-1a5) and the other is centered around the antinodal ($\mathit{\pi}$,0) region (Fig. 1b1-1b5). Each Fermi surface mapping is obtained by using our ARToF analyzer which can simultaneously cover two-dimensional momentum space with high energy and momentum resolutions. The entire Fermi surface of Bi2201 is measured by combining the nodal and antinodal Fermi surface mappings in Fig. 1a1-1a5 and Fig. 1b1-1b5. It is well-known that, in Bi-based cuprate superconductors, the structural modulations along the $\Gamma$-$\mathrm{Y}$ direction give rise to superstructure bands, i.e., extra replica bands that are formed by
    shifting the original Fermi surface by $\pm n\mathbf{Q}$, where $\mathbf{Q}$ is the
    vector of the structural modulation and $n$ is the order of the superstructure bands\cite{KKadowaki1994PAebi,GJennings1995HDing,KKadowaki1995JOsterwalder,XJZhou2019SJLiu,XJZhou2020QGao}. In addition, there are also shadow bands and the superstructure bands of the shadow bands\cite{KKadowaki1994PAebi,GJennings1995HDing,KKadowaki1995JOsterwalder}. As depicted in Fig. 1a5 and Fig. 1b5, all the observed Fermi surface sheets can be well assigned to the main Fermi surface (MB, thick red line), the first-order superstructure bands of the main Fermi surface (SSB$\_$1, solid pink line), the second-order superstructure bands of the main Fermi surface (SSB$\_$2, dashed pink line), the shadow band of the  main Fermi surface (SDB, purple line) and the first-order superstructure bands of the SDB shadow band (SDB$\_$SS1, blue line). The main Fermi surface stands out clearly in all the measurements (thick red lines in Fig. 1) although it is complicated by other Fermi surface sheets, particularly near the antinodal region. The main Fermi surface exhibits little change with temperature over the whole temperature range of 20$\sim$200\,K, as seen in Fig. 1 where the same thick red lines agree well with the observed main Fermi surface at different temperatures.
    
	Figure 2 shows the temperature dependence of the band structures in the Opt32K Bi2201 sample measured along three typical momentum cuts near the antinodal region. In order to directly visualize the gap opening and the particle-hole symmetry, the presented band structures in Fig. 2a1-2c5 are obtained by dividing the original data with the corresponding Fermi-Dirac distribution functions to show the electronic states above the Fermi level. The corresponding photoemission spectra (energy distribution curves, EDCs) are presented in Fig. 3. To better understand the data, we simulated the single-particle spectral function of a conventional BCS superconductor in the normal state (Fig. 2f) and in the superconducting state (Fig. 2g). In this case, the particle-hole symmetry is conserved which can be judged from two aspects. The first is that the Fermi momentum $\mathbf{k}_\mathrm{F}$ keeps fixed in the normal and superconducting state. The second is that the single-particle spectral function $A(\mathbf{k},\omega)$ satisfies $A(\mathbf{k}_\mathrm{F},\omega)$=$A(\mathbf{k}_\mathrm{F},-\omega)$ at the Fermi momentum $\mathbf{k}_\mathrm{F}$. The gap opening corresponds to the spectral weight suppression at the Fermi level. 
	
	We find that the particle-hole symmetry is conserved in both the pseudogap state and the superconducting state near the antinodal region as seen in Fig. 2 and Fig. 3. First, the Fermi momentum shows little change upon crossing the pseudogap transition and the superconducting transition. Fig. 2d shows the momentum distribution curves (MDCs) at the Fermi level obtained from the band structures measured along the momentum cut 2 at different temperatures (Fig. 2b1-2b5). No obvious change of the two Fermi momenta ($\mathbf{k}_\mathrm{FL}$ and $\mathbf{k}_\mathrm{FR}$) is observed in the measured temperature range of 20$\sim$200\,K. The same is true for the Fermi momentum from the antinodal cut 3 that is plotted in Fig. 2e. These are consistent with the fixed Fermi surface observed at different temperatures in Fig. 1. Second, as the temperature decreases from 200\,K, the pseudogap opening between $T^{*}$ and $T_{c}$ and the superconducting gap opening below $T_{c}$ can be directly visualized from the spectral weight suppression at the Fermi level in the measured band structures shown in Fig. 2a1-2c5. One can also see from these band structures that, when either the pseudogap or the superconducting gap opens, the spectral function at the Fermi momentum is nearly symmetric with respect to the Fermi level. These can be directly observed from EDCs at the Fermi momentum shown in Fig. 3 (blue and red curves) which are nearly symmetric with respect to the Fermi level.    
	
	Our measured results of the optimally-doped La-Bi2201 near the antinodal region shown in Fig. 1-Fig. 3 are rather different from the previous reports on the optimally-doped $\mathrm{Pb_{0.55}Bi_{1.5}Sr_{1.6}La_{0.4}CuO_{6+\delta}}$\cite{ZXShen2010MHashimoto,ZXShen2011RHHe}. In that case, dramatic electronic structure changes are observed both across $T^{*}$ and across $T_{c}$ near the antinodal region. The Fermi momentum and the corresponding Fermi surface exhibit an obvious change across the pseudogap transition $T^{*}$ indicating the breaking of the particle-hole symmetry (also plotted in Fig. 2e for comparison)\cite{ZXShen2010MHashimoto,ZXShen2011RHHe}. A complex structure with two energy scales below the Fermi level develops below $T_{c}$ in the superconducting state which can not be derived by the BCS formula from the band structure in the normal state above $T^{*}$. In our case, except for the narrow energy range near the Fermi level which is sensitive to the gap opening, most of the band structures do not show obvious change with temperature in the entire range of 20$\sim$200\,K, as seen in Fig. 2. The Fermi momentum and the corresponding Fermi surface do not change across $T^{*}$ and the particle-hole symmetry is observed in the pseudogap state. In the superconducting state, we do not observe the complex electronic structures reported before\cite{ZXShen2011RHHe} and the electronic structures in the superconducting state can be well connected to the normal state by the BCS picture. The origin of the big difference between our results and the previous measurements\cite{ZXShen2010MHashimoto,ZXShen2011RHHe} needs to be further investigated. We note that, in the optimally-doped Bi2212 superconductor, no obvious electronic structure changes are observed across $T^{*}$ and $T_{c}$ and the particle-hole symmetry is conserved in the pseudogap state and the superconducting state\cite{XJZhou2018XSun}. These observations are consistent with our present results on Bi2201. They indicate that the particle-hole symmetry breaking across $T^{*}$ reported before\cite{ZXShen2010MHashimoto,ZXShen2011RHHe} is not universal in high temperature cuprate superconductors.   
	
	Now we come to examine the momentum dependence of the particle-hole symmetry along the Fermi surface. In the underdoped Bi2212, it is reported that, in the pseudogap state, the particle-hole symmetry is conserved near the antinodal region but breaks near the nodal region\cite{GDGu2008HBYang}. To this end, we show EDCs along the whole Fermi surface measured at different temperatures in Fig. 4. We find that, when the pseudogap develops between $T^{*}$ and $T_{c}$, or the superconducting gap opens below $T_{c}$, all the EDCs along the Fermi surface are nearly symmetric with respect to the Fermi level. These results indicate that the particle-hole symmetry is conserved along the entire Fermi surface both in the pseudogap state and in the superconducting state. 
	
	In summary, by taking high-resolution laser-based ARPES measurements on the optimally-doped Bi2201, we have observed the particle-hole symmetry conservation across the pseudogap transition along the entire Fermi surface. These results provide key information to understand the nature of the pseudogap and its relation with high temperature superconductivity in cuprate superconductors.  
    
    \vspace{3mm}
    

    \vspace{3mm}
    
    \noindent {\bf Acknowledgement} This work is supported by the National Natural Science Foundation of China (Grant Nos. 11888101, 11922414 and 11974404), the National Key Research and Development Program of China (Grant Nos. 2021YFA1401800, 2017YFA0302900, 2018YFA0305602 and 2018YFA0704200), the Strategic Priority Research Program (B) of the Chinese Academy of Sciences (Grant No. XDB25000000 and XDB33000000), the Youth Innovation Promotion Association of CAS (Grant No. 2021006), the Synergetic Extreme Condition User Facility (SECUF) and the Research Program of Beijing Academy of Quantum Information Sciences (Grant No. Y18G06).

    \vspace{3mm}
    
    \noindent {\bf Author Contributions}\\
    X.J.Z. and H.T.Y. proposed and designed the research. H.T.Y. carried out the ARPES experiments. H.T.Y. grew the single crystals. H.T.Y., Q.G., C.Y.S., C.H.Y., Y.W.C., F.F.Z., F.Y., S.J.Z., Q.J.P., G.D.L., L.Z., Z.Y.X. and X.J.Z. contributed to the development and maintenance of Laser-ARPES systems. H.T.Y and X.J.Z. analyzed the data and wrote the paper. All authors participated in discussions and comments on the paper.
       
    \vspace{3mm}
    
    \noindent {\bf\large Additional information}\\
    \noindent{\bf Competing financial interests:} The authors declare no competing financial interests.

    \newpage
    
    \begin{figure*}[tpb]
    \begin{center}
    	\includegraphics[width=1.0\columnwidth,angle=0]{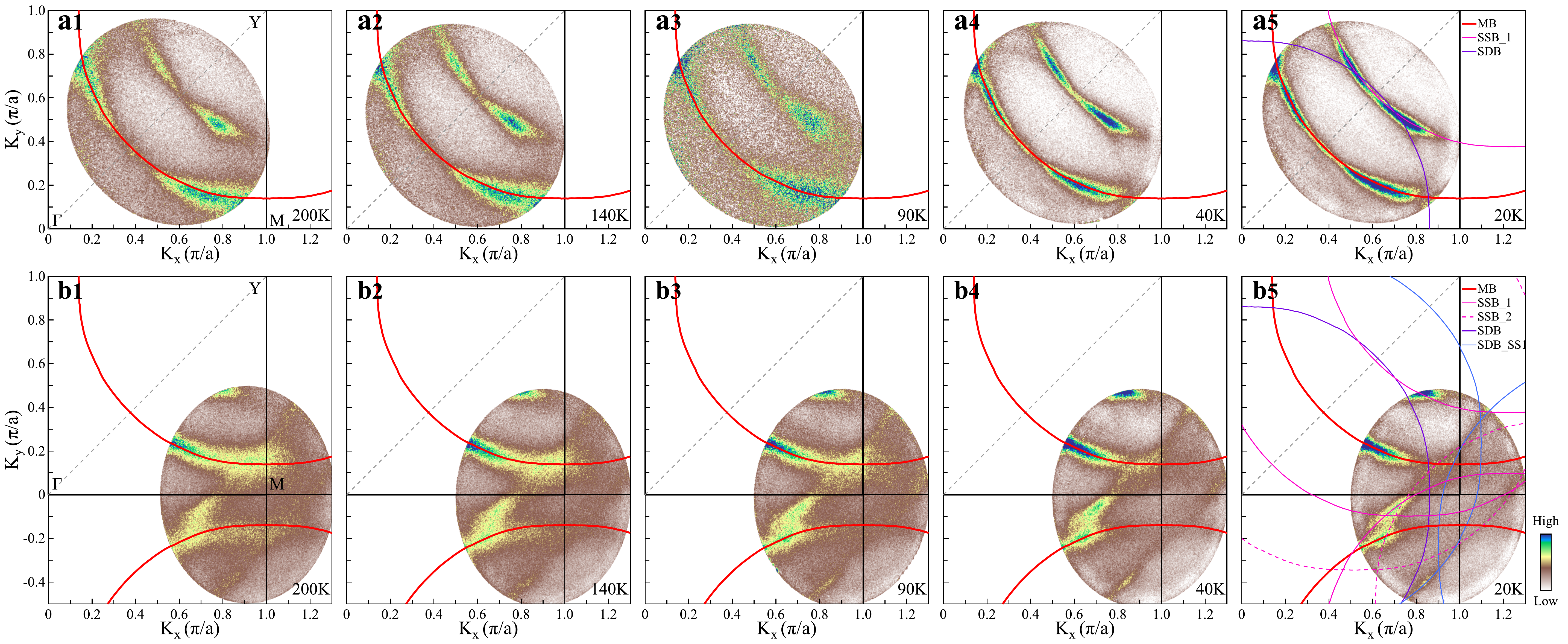}
    \end{center}
    
    \caption{{\bf Fermi surface mappings of Opt32K Bi2201 measured at different temperatures.} (a1-a5) Fermi surface mappings near the nodal region measured at different temperatures of 200\,K (a1), 140\,K (a2), 90\,K (a3), 40\,K (a4) and 20\,K (a5). The measured main Fermi surface is marked by the red lines. The pink line (SSB$\_$1) in (a5) represents the first-order superstructure band of the main Fermi surface. The purple line (SDB) in (a5) indicates the shadow band of the main Fermi surface. (b1-b5) Same as (a1-a5) but measured near the antinodal region. The red line (MB) shows the main Fermi surface. In (b5), the pink solid line (SSB$\_$1) and pink dashed line (SSB$\_$2) represent the first-order and second-order superstructure bands of the main Fermi surface, respectively; the purple line (SDB) represents the shadow band of the main Fermi surface; the blue line (SDB$\_$SS1) represents the first-order superstructure band of the SDB shadow band.}
   
    \end{figure*}

    \begin{figure*}[tbp]
	\begin{center}
		\includegraphics[width=1.0\columnwidth,angle=0]{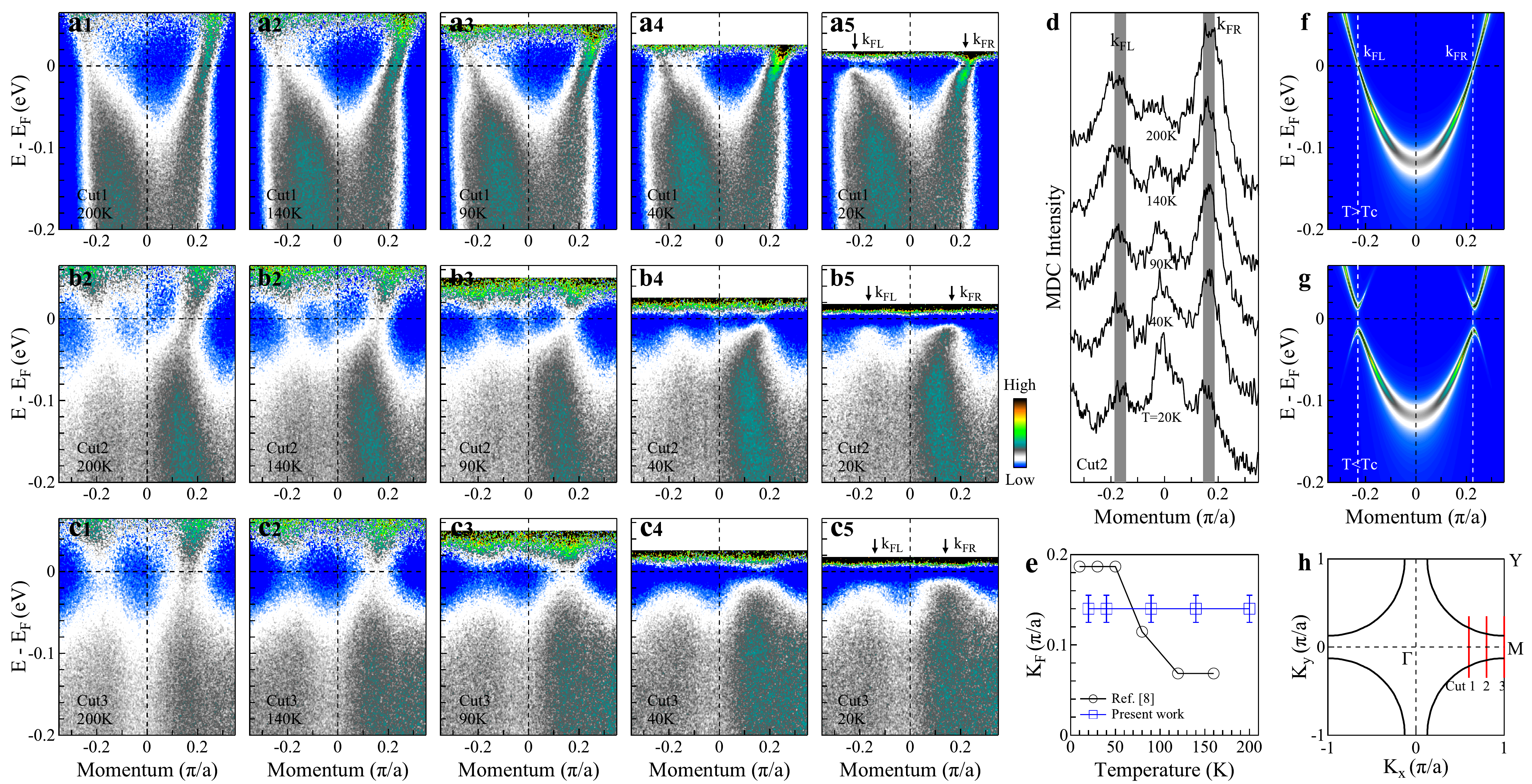}
	\end{center}
	
	 \caption{{\bf Temperature dependence of the band structures near the antinodal region in Opt32K Bi2201.} (a1-a5) Band structures along the momentum cut 1 measured  at different temperatures of 200\,K (a1), 140\,K (a2), 90\,K (a3), 40\,K (a4) and 20\,K (a5). The location of the momentum cut 1 is shown in (h) by a red line. The Fermi-Dirac distribution function is removed from the images. The Fermi momenta are marked by two arrows labeled as $\mathrm{k_{FL}}$ and $\mathrm{k_{FR}}$ in (a5).  (b1-b5) Same as (a1-a5) but measured along the momentum cut 2. (c1-c5) Same as (a1-a5) but measured along the momentum cut 3. (d) Momentum distribution curves (MDCs) at the Fermi level obtained from (b1-b5). The two main peaks are marked as $\mathrm{k_{FL}}$ and $\mathrm{k_{FR}}$, corresponding to the Fermi momenta in (b1-b5). For clarity, the data are offset along the vertical axis. (e) Fermi momentum at different temperatures (blue empty squares) obtained from the antinodal cut 3 measurements. For comparison, the antinodal Fermi momentum change with temperature from the previous measurements\cite{ZXShen2010MHashimoto,ZXShen2011RHHe} is also plotted (black empty circles). (f) The simulated single-particle spectral function in the normal state. (g) The corresponding single-particle spectral function in the superconducting state simulated by using the BCS formula\cite{JRSchrieffer1957JBardeen}. The superconducting gap is 15\,meV used in the simulation. (h) Schematic Fermi surface of the Opt32K Bi2201 and the location of the momentum cuts.}

	\end{figure*}

    \begin{figure*}[tbp]
	\begin{center}
		\includegraphics[width=1.0\columnwidth,angle=0]{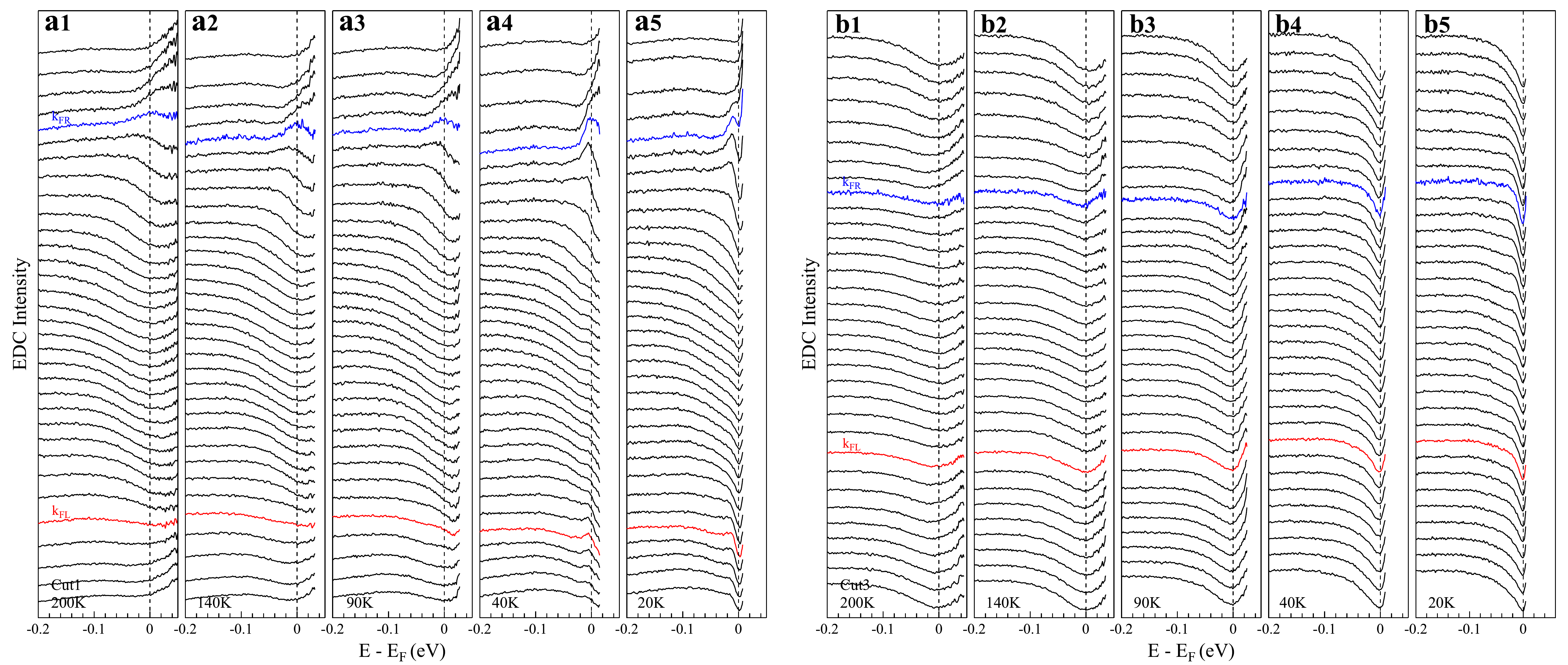}
	\end{center}
	
	\caption{{\bf Temperature dependence of the energy distribution curves (EDCs) in Opt32K Bi2201 measured along two typical momentum cuts.} (a1-a5) EDCs along the momentum cut 1 measured at different temperatures of 200\,K (a1), 140\,K (a2), 90\,K (a3), 40\,K (a4) and 20\,K (a5) obtained from the images in Fig. 2(a1-a5). The EDCs at the Fermi momenta, $\mathrm{k_{FR}}$ and $\mathrm{k_{FL}}$, are highlighted by the blue and red lines, respectively. For clarity, the EDCs are offset along the vertical axis. (b1-b5) Same as (a1-a5) but for the momentum cut 3 obtained from the images in Fig. 2(c1-c5).} 
	
	\end{figure*}

    \begin{figure*}[tbp]
	\begin{center}
		\includegraphics[width=1.0\columnwidth,angle=0]{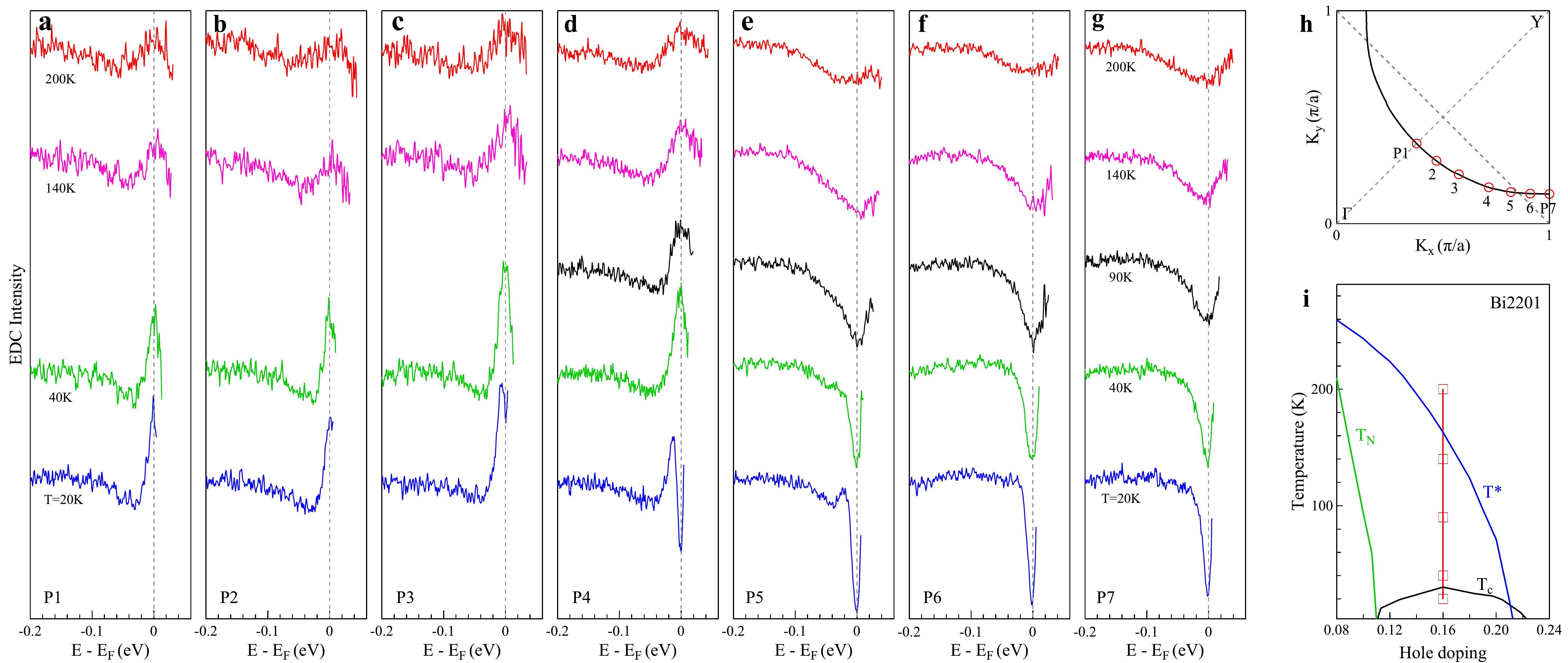}
	\end{center}
	
	\caption{{\bf Particle-hole symmetry along the entire Fermi surface of  Opt32K Bi2201 in the pseudogap state and in the superconducting state.} (a) EDCs measured at the Fermi momentum P1 at different temperatures. The location of the P1 point is indicated in (h). The Fermi-Dirac distribution function is removed in the EDCs. For clarity, the EDCs are offset along the vertical axis. (b-g) Same as (a) but measured at the momentum points of 2 (b), 3 (c), 4 (d), 5 (e), 6 (f) and 7 (g). (h) Schematic Fermi surface of Bi2201 and the location of the momentum points P1-P7 along the Fermi surface. (i) Schematic phase diagram of Bi2201\cite{CTLin2005GQZheng}. The blue, black and green lines show the temperatures of pseudogap, superconductivity and antiferromagnetic order. The red line indicates the temperature range of our ARPES measurements.}

	\end{figure*}

\end{document}